\documentclass[aps,prd,longbibliography,onecolumn,showpacs,amsmath,amssymb,superscriptaddress,nofootinbib,floatfix,preprintnumbers,secnumarabic]{revtex4-1}

\usepackage{graphicx}
\usepackage[dvipsnames]{xcolor}
\newcommand{\remove}[1]{}
\usepackage{xspace}
\definecolor{darkblue}{rgb}{0,0,0.5}
\definecolor{darkgreen}{rgb}{0.1,0,0.3}
\definecolor{darkred}{rgb}{0.6,0,0}
\usepackage[bookmarks=false, bookmarksnumbered=true, colorlinks=true, urlcolor=darkblue, citecolor=darkgreen, linkcolor=darkred, breaklinks=true]{hyperref} 
\usepackage{siunitx}
\usepackage[normalem]{ulem}
\usepackage[capitalise]{cleveref}
\usepackage{multirow}

\newcommand{\mueff}{\mu^*}

\newcommand{\coherent}{CE$\nu$NS\xspace}
\newcommand{\ubl}{$U(1)_{B-L}$ \xspace}

\usepackage{fullpage}
\usepackage{geometry}      
\begin{document}

\preprint{IFT-UAM/CSIC-21-71}
\preprint{CP3-21-38 }
\preprint{IPPP/20/119}

\vspace*{0.7cm}

\title{Medium effects in supernovae constraints on light mediators}  

\author{David G. Cerde\~no}
\email{davidg.cerdeno@gmail.com}
\affiliation{Instituto de F\' isica Te\'orica, IFT-UAM/CSIC, 28049 Madrid, Spain}
\affiliation{Departamento de F\' isica Te\'orica, Universidad Aut\'onoma de Madrid, 28049 Madrid, Spain}

\author{Marina Cerme\~no}
\email{marina.cermeno@uclouvain.be}
\affiliation{Centre for Cosmology, Particle Physics and Phenomenology (CP3),Universit\'e catholique de Louvain, Chemin du Cyclotron 2,B-1348 Louvain-la-Neuve, Belgium} 

\author{M. \'Angeles P\'erez-Garc\'\i a}
\email{mperezga@usal.es}
\affiliation{Department of Fundamental Physics, University of Salamanca, Plaza de la Merced s/n 37008 Salamanca, Spain}

\author{Elliott Reid}
\email{elliott.m.reid@durham.ac.uk}
\affiliation{Institute for Particle Physics Phenomenology, Durham University, Durham DH1 3LE, United Kingdom}

\begin{abstract}
In this article, we reevaluate supernovae (SN) constraints on the diffusion time of neutrinos for a family of extensions of the Standard Model that incorporate new light scalar and vector mediators. We compute the neutrino mean free path, taking into account medium effects in the neutrino-nucleon scattering cross-section, and a radial dependence of the density, energy, and temperature inside the proto-neutron star to determine the coupling strengths compatible with SN1987A constraints on the time duration signal of diffusing neutrinos. 
We show that medium effects can induce an order of magnitude enhancement in the neutrino mean free path with respect to the vacuum calculation. The increase is more significant when new physics terms dominate over the Standard Model contribution (that is, for small mediator mass and large couplings).
Finally, we interpret these results as bounds on the parameter space of a vector \ubl model and scalar lepton number conserving and lepton number violating scenarios, improving on previous results in the literature where medium effects were ignored.
We show that SN constraints on the neutrino diffusion time lie within regions of the parameter space that are already ruled out by other experimental constraints. 
We also comment on potential limits due to changes in the SN equation of state or right-handed neutrino free-streaming, but argue that detailed numerical simulations are needed to improve the reliability of these limits.
\end{abstract}
\maketitle

\section{Introduction}
\label{sec:Intro}

There is growing interest in particle physics constructions that feature new low-mass mediators as an extension of the Standard Model (SM). This interest has been fuelled in part by observed anomalies in low-energy observables, such as the long-standing discrepancy in the muon anomalous magnetic moment, and in part by the potential for these models to connect to an otherwise secluded sector that could account for the dark matter content of the universe. In the absence of any clear signal of new physics either from the high-energy regime explored at the LHC or from direct detection searches for canonical weakly-interacting massive particles, alternative search strategies are gaining attention, including those that favour low-energy precision measurements across a wide range of experimental techniques.

In general, particle models with light mediators include new interactions in the neutrino sector, which can often be interpreted in terms of an effective theory and described as non-standard interactions (NSI). This is a particularly interesting possibility, as it widens the range of experimental probes of these constructions. For example, data from dedicated neutrino detectors can be used to set constraints on the resulting neutrino-electron scattering, as no deviation has yet been found from the SM prediction. Experimental data from  GEMMA \cite{Beda:2013mta} and Borexino \cite{Agostini:2017ixy} lead to competitive bounds on the parameter space of these models for mediator masses of the order of the MeV \cite{Khan:2019jvr,Amaral:2020tga}. 
Likewise, new physics in the neutrino sector can also be probed through coherent elastic neutrino nucleus scattering (\coherent). This rare process, which in the SM occurs at leading-order through the exchange of a $Z$ boson \cite{Freedman:1973yd}, has been recently observed by the COHERENT collaboration, employing two different targets. Their original results on CsI \cite{Akimov:2017ade} and the latter measurement in liquid argon \cite{Akimov:2020pdx} are in agreement with the SM prediction, which allows to set limits on the parameter space of new-physics scenarios with light mediators \cite{Abdullah:2018ykz,Banerjee:2018eaf,Papoulias:2019txv,Khan:2019cvi, Miranda:2020tif,Amaral:2020tga,Amaral:2021rzw}. The large exposure of projected future experiments at spallation neutrino sources \cite{Baxter:2019mcx,ccm,Akimov:2019xdj} makes them well suited to probe mediator masses in the MeV range.

Dark matter direct detection experiments have recently joined the quest for light mediators. Although they were not specifically designed for this aim, these detectors can be sensitive to neutrino interactions with the electrons and nuclei of their target. In fact, due to the resemblance of this signal with that expected from a weakly-interacting massive particle, \coherent is considered a background in the search for dark matter particles \cite{Billard:2013qya}, and it can be interpreted as a 'neutrino floor' in the dark matter-nucleon scattering cross section \cite{Vergados:2008jp}. New physics in the neutrino sector can alter the SM predictions for nuclear recoils \cite{Cerdeno:2016sfi,Dent:2016wor}, thereby raising the neutrino floor especially at low dark matter masses \cite{Boehm:2018sux,Sadhukhan:2020etu}. Likewise, the effects of new physics could also be felt in the electron recoil spectrum. So far, the absence of a confirmed signal in either electron and nuclear recoils has been translated into constraints on simplified models and scenarios with non-standard neutrino interactions \cite{Cerdeno:2016sfi,Essig:2018tss,Gonzalez-Garcia:2018dep, Amaral:2020tga,Khan:2020csx,Boehm:2020ltd}.

The presence of light new states can also have crucial implications in cosmological and astrophysical observations. In particular, a bound on the effective number of relativistic degrees of freedom can be derived from Big Bang nucleosynthesis, that sets stringent constraints on the abundance of new particles with masses below 1~MeV \cite{Huang:2017egl}.

Finally, light particles can be produced in stellar interiors and alter different aspects of stellar evolution, in particular, SN explosions. The observed neutrino flux from SN1987A \cite{Bionta:1987qt, Hirata:1987hu, Alekseev:1987ej} permitted the study of neutrino properties (setting upper limits on their mass and lifetime) and set bounds on new physics in this sector \cite{Sato:1987rd, Spergel:1987ch,Bahcall:1987ua,Burrows:1987zz,Schramm:1990pf,Loredo:2001rx,Frieman:1987as,Kolb:1988pe,Berezhiani:1989za,Farzan:2002wx,Heurtier:2016otg, Farzan:2018gtr, Suliga:2020jfa}. Most of these constraints address the cooling effect of light weakly-interacting particles, which would escape the star unimpeded (thus leading to a different neutrino luminosity than the one actually observed). This is applicable to a wide range of models that include axions (or axion-like particles) as the most representative example. It also affects new mediators in the neutrino sector, for example light gauge bosons (dark photons) \cite{Dent:2012mx,Rrapaj:2015wgs,Mahoney:2017jqk,Chang:2016ntp,Hardy:2016kme} or scalar particles \cite{Krnjaic:2015mbs,Dev:2020eam}. It should be noted that cooling constraints generally restrict a window of very small couplings (as the particle needs to escape the star and these bounds do not apply if it is reabsorbed after production), however, large couplings can affect the diffusive properties of neutrinos \cite{reddy}, decreasing their mean free-path and leading to an emitted flux which can exceed the observed $\sim10$~s. This has been used in Refs.~\cite{Farzan:2018gtr, Suliga:2020jfa} to derive upper bounds in the neutrino coupling to new scalar and vector mediators, showing that they can compete with limits from other experimental sources. As it was then shown in Ref.~\cite{Boehm:2018sux}, these constraints are relevant, since they determine the maximum height of the neutrino floor in direct dark matter detection experiments.

In this article, we carry out an improved computation of the neutrino mean free path to derive a more reliable upper limit on the mediator coupling to neutrinos. To do this, our analysis takes into account the radial dependence of the density, energy, and temperature in the proto-neutron star (proto-NS). More importantly, given that the density can be several times the nuclear saturation density, $\rho_0 \sim  2 \times 10^{14} \; \rm g \; cm^{-3}$, we include medium effects in the computation of the neutrino-nucleon scattering cross-section, which were not included in the analysis of Refs.\,\cite{Farzan:2018gtr, Suliga:2020jfa}. Medium effects effects can increase the neutrino mean free path by approximately an order of magnitude  \cite{reddy, Burrows:1998cg, Burrows:2002jv, Roberts:2012um, Roberts:2016mwj} with pure SM interactions, and in this work we extend the analysis to new physics, showing that they can be even more important for low mass mediators. This can significantly alter the resulting neutrino diffusion time. We interpret our results as upper bounds on the neutrino couplings in simplified models, for which we choose a (lepton number violating and lepton number conserving) scalar and a \ubl vector model.

\vspace*{1.5ex}

This article is organised as follows. In \cref{sec:models} we introduce simplified models that incorporate a new vector or scalar coupling to the SM and provide the new matrix elements that contribute to neutrino-nucleon scattering. In \cref{sec:SN} we review the main constraints on light new mediators that can be derived from SN physics. We explain in detail how to compute the neutrino mean free path, for which we consider medium effects and a radial dependence of the density and temperature.  In \cref{sec:discussion} we show the medium effects on the neutrino mean free path and study the area of the two-dimensional parameter space that is excluded by SN constraints, putting it in context with bounds from other detection methods.
Finally, our conclusions are presented in \cref{sec:conclusions}.

\section{Simplified models of new physics with light mediators}
\label{sec:models}

\begin{figure}[t!]
    \centering
    \includegraphics[width = 0.35\textwidth]{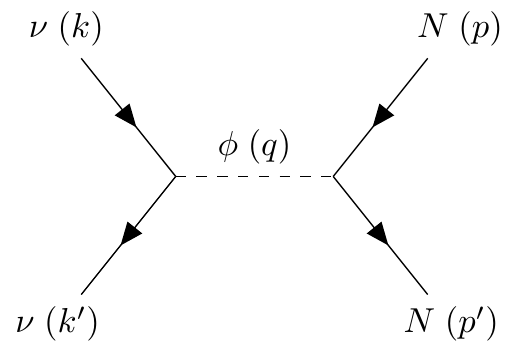}\hspace*{2cm}
    \includegraphics[width = 0.35\textwidth]{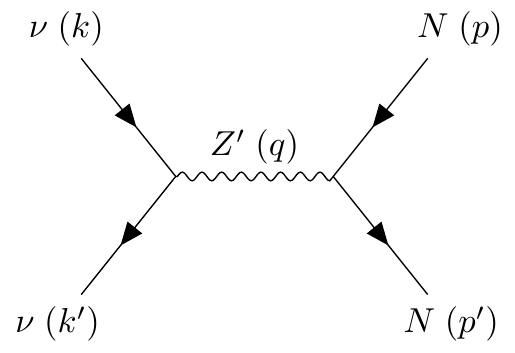}
    \caption{Contributions to neutrino-nucleon scattering for a scalar mediator (left) and a vector mediator (right). }
    \label{fig:feynman}
\end{figure}

We consider two classes of low-scale simplified models, which introduce a new light mediator (scalar or vector) that provides extra interactions between neutrinos and SM fermions (quarks and/or leptons)~\cite{Boehm:2004uq}. We do not address here how to realise these scenarios as complete UV models, as this might imply adding extra fields for anomaly cancellation. The new interactions induce corrections to neutrino-nucleon scattering as illustrated in \cref{fig:feynman}, thereby altering the neutrino diffusion time. In this section we compute the new matrix elements that enter the computation of the neutrino mean free path.

In the following, we denote $p'^{\mu}=(E'_N,\vec{p'})$ and  $p^{\mu}=(E_N,\vec{p})$ as the four-momentum for the outgoing and incoming nucleon of effective mass $m_N^*$, respectively, and $k'^{\mu}=(E'_\nu,\vec{k'})$ and $k^{\mu}=(E_\nu,\vec{k})$ the analogous for neutrinos of mass $m_\nu$. The transferred four-momentum is defined as $q^\mu=(q_0, \vec{q})=p'^{\mu}-p^{\mu}=k^{\mu}-k'^{\mu}$. We will give more details about the concept of effective mass in Section III.

\subsection{Scalar mediator}
\label{sec:scalar}

As a first example, we introduce a new light scalar field, $\phi$, of mass $m_\phi$ that interacts with SM fermions. The scalar must necessarily couple a left handed fermion field to a right handed one, which can lead to two different scenarios for neutrinos, which we study separately.

First, we can consider a lepton number violating (LNV) scenario with Majorana neutrinos. In this case, we assume that the masses of possible heavy neutrino states are much heavier than the typical energies inside the supernova and their interactions can be disregarded.  Alternatively, assuming Dirac neutrinos, the scalar field might couple to right handed neutrinos in a lepton number conserving (LNC) model. In this case, different regimes are possible for the right handed neutrino mass, $m_{\nu_R}$. In this work we consider that $m_{\nu_R} \lesssim {\cal O}(1\,{\rm eV})$, and therefore the relevant cross sections for neutrino scattering mediated by the new LNC scalar in most terrestrial experiments are nearly identical to the LNV case\footnote{If $m_{\nu_R} \gtrsim {\cal O}(1\,{\rm GeV})$, the neutrino-scattering cross section would be kinematically suppressed when the centre-of-mass energy of the collision is smaller than the $\nu_R$ mass, and right handed neutrinos would not be produced in any of the scenarios that we consider. For the intermediate case where the $\nu_R$ mass is of the order of the MeV, scattering would still occur at some or all of the energy scales we consider, with the presence of the massive $\nu_R$ producing potentially measurable effects that could even allow us to determine $m_{\nu_R}$. This would produce a complex phenomenology which, while interesting, is beyond the scope of this work.}.

In the case of an LNV scalar, the new interaction terms in the Lagrangian take the form
\begin{equation}\label{eq:LNV_Lagrangian}
    \mathcal{L_{\mathrm{LNV}}} \supset -C_v \bar{\nu}_L^c \phi \nu_L - \sum_q C_q \bar{q} \phi q - \sum_l C_l \bar{l} \phi l,
\end{equation}
where $C_v$, $C_q$, $C_l$ and are the couplings to neutrinos, quarks, and leptons, respectively. This differs from the LNC scalar only in the first term, with 
\begin{equation}\label{eq:LNC_Lagrangian}
    \mathcal{L_{\mathrm{LNC}}} \supset -C_v \bar{\nu}_R \phi \nu_L - \sum_q C_q \bar{q} \phi q - \sum_l C_l \bar{l} \phi l.
\end{equation}
Within SN, we consider neutrinos scattering coherently with nucleons. The quark contribution to the Lagrangian can therefore be rewritten in terms of whole nucleons as
\begin{equation}
    \mathcal{L}_{N \phi}=\bar{\psi}_N C_N \psi_N \phi,
\end{equation}
with
\begin{equation}
    C_N=m_N \left[\sum_{q} C_q \frac{f_q^N}{m_q} \right],
\end{equation}
where values of $m_q$ and $f_q^N$ can be found in Ref. \cite{Farzan:2018gtr}.
For universal $Q_q$, as considered in our work, one obtains $C_N = 13.8\, C_q$. 
The presence of the additional scalar provides a new channel for neutrino-nucleon scattering. The squared matrix element 
is modified from its SM value by an additional term
\begin{equation}
    |\overline{\mathcal{M}}|^2 = 
    |\overline{\mathcal{M}}|_{SM}^2 + |\overline{\mathcal{M}}|_{S}^2\ ,
\end{equation}
with
\begin{equation}
    \label{eq:Msq_scalar}
    |\mathcal{\overline{M}}|_S^2=\frac{- C_\nu^2 C_N^2}{(q^2-m_\phi^2)^2} q^2(4m_N^{*2}-q^2).
\end{equation}

In principle, $C_\nu$ and $C_N$ are separate free parameters in our model. We can reduce the dimensions of the parameter space by placing constraints on the effective coupling $Y=\sqrt{C_\nu C_N}$. Any other process that involves neutrino-quark scattering can directly set bounds on $Y$, and so can be immediately compared with our constraints from SN physics. Limits on the individual couplings $C_\nu$ and $C_q$ can also be combined to obtain a constraint on $Y$. However, applying bounds from other processes require us to make assumptions about the relation between the couplings to different particles.

Constraints from neutrino scattering (either \coherent or neutrino-electron scattering) in terrestrial experiments are practically identical for both a LNC and a LNV scalar (with $m_{\nu_R}\lesssim 1 \; \rm eV$). However, each species of scalar has its own unique constraints as well, stemming from limits on either the production of right handed neutrinos, or the rate of LNV processes. Furthermore, the rate of neutrino scattering within the dense interior of the SN differ for LNV and LNC scalars due to the different nature  (Majorana and Dirac, respectively) of neutrinos and to the fact that LNV interactions induce changes in the chemical potentials of the particles in the medium, as it will be discussed in \cref{sec:diff}.

\subsection{Vector mediator: Gauged $U(1)_{B-L}$}
\label{sec:b-l}

As a paradigm of vector-mediated scenarios, we consider a \ubl model. The symmetry is spontaneously broken, leading to a new vector mediator, $Z'$, with mass $m_{Z'}$, which, depending on the scale of the symmetry breaking, can be much lighter than the SM $Z$ boson.

The new interaction terms in the Lagrangian take the form
\begin{equation}\label{eq:BL_Lagrangian}
    \mathcal{L_{\mathrm{B-L}}} = - \sum_q C_q \bar{q} \gamma^\mu q Z^\prime_\mu - \sum_l C_l \bar{l} \gamma^\mu l Z^\prime_\mu - \sum_\nu C_\nu \bar{\nu} \gamma^\mu \nu Z^\prime_\mu ,
\end{equation}
where all quarks have the same charge, $C_q=g_{B-L}/3$, and all charged and neutral leptons have charge $C_{l,\nu}=-g_{B-L}$. The $Z'$ may also interact with SM fermions via kinetic mixing with either the SM $Z$ or the photon. Here we assume that the couplings to SM fermions induced through kinetic mixing are subdominant to the direct couplings $C_i$, and so we neglect kinetic mixing effects.

Gauging the \ubl symmetry in a theory with only SM fermions leads to non-zero anomalies, so additional particles must be added to ensure these anomalies are cancelled out. Although several non-minimal solutions exist, the simplest way is to include three right handed neutrinos \cite{Nanda:2017bmi}. In this case the anomalies from the left and right handed neutrino loops exactly cancel and the theory is anomaly free. As in the case of the LNC scalar considered in the previous section, neutrinos in this model are Dirac fermions and interactions with the $U(1)_{B-L}$ mediator do not violate lepton number. There is thus no mechanism for the efficient conversion of electron neutrinos to neutrinos and antineutrinos of other flavours which can deleptonize the star or impact properties such as the entropy or the temperature of the SN core. Unlike in that model, right handed neutrinos are not produced in t-channel neutrino scattering events mediated by a \ubl mediator. We therefore assume that right handed neutrinos are heavy enough so as not to be produced inside the SN (this requires masses above the GeV scale, natural in most see-saw models). This choice has no effect on the \coherent\ or neutrino-electron scattering cross sections in the \ubl model.

Considering this scenario, the total matrix element contains an interference term between $Z'$ and the SM $Z$ boson, 
\begin{equation}
    |\overline{\mathcal{M}}|^2 = |\overline{\mathcal{M}}|_{SM}^2 + |\overline{\mathcal{M}}|_{BL}^2 + \frac{1}{2}\sum_{\rm spins} 2 \mathrm{Re}\left( \mathcal{M}_{BL}\mathcal{M}^{*}_{SM} \right)\ .
\end{equation}
Since neutrinos are Dirac, only left handed neutrinos and right handed antineutrinos couple to the $Z$ and $Z'$ vector mediators. Remember also that we are only concerned with the light neutrino states, as the heavy ones are assumed to have masses well above the GeV and are therefore not produced in the SN.

Both for the left handed neutrino nucleon interaction, $\nu N \rightarrow \nu N$, and for the right handed antineutrino nucleon one, $\bar{\nu} N \rightarrow \bar{\nu} N$, the pure BL term reads
\begin{equation}
    |\mathcal{\overline{M}}|^2_{BL}=
    \frac{2 g_{BL}^4}{(q^2-m_{Z'}^2)^2}
    \left(8 (pk)^2+q^2(2m_N^{*2}+q^2+4pk) \right).
    \label{eq:MBL}
\end{equation}

On the other hand, as the SM amplitude is different for $\nu N \rightarrow \nu N$ and $\bar{\nu} N \rightarrow \bar{\nu} N$, the mixing term is different too.
For $\nu N \rightarrow \nu N$ it reads
\begin{align}
    \sum_{\rm spins}  \mathrm{Re}\left( \mathcal{M}_{BL}\mathcal{M}^{\nu*}_{SM} \right) =&\frac{-4\sqrt{2}g_{BL}^2G_F}{q^2-m_{Z'}^2}  \left(q^2(\tilde{F}_1^{N }(q^2)+ \tilde{F}_A^{N }(q^2))(q^2+2(m_N^{*2}+2pk))
    \nonumber\right.\\ 
    & \left.+8 (pk)^2 \tilde{F}_1^{N }(q^2)-2\tilde{F}_A^{N }(q^2) m_N^{*2}q^2\right),
    \label{eq:Mmixingnu}
\end{align}
and for $\bar{\nu} N \rightarrow \bar{\nu} N$
\begin{align}
    \sum_{\rm spins}  \mathrm{Re}\left( \mathcal{M}_{BL}\mathcal{M}^{\bar{\nu}*}_{SM} \right) =&\frac{-4\sqrt{2}g_{BL}^2G_F}{q^2-m_{Z'}^2}  \left(q^2(\tilde{F}_1^{N }(q^2)- \tilde{F}_A^{N }(q^2))(q^2+2(m_N^{*2}+2pk))
    \nonumber\right.\\ 
    & \left.+8 (pk)^2 \tilde{F}_1^{N }(q^2)+2\tilde{F}_A^{N }(q^2) m_N^{*2}q^2\right).
    \label{eq:Mmixingantinu}
\end{align}

In these expressions, $\tilde{F}_1^{N }(q^2)$ is the neutral current vector form factor and $\tilde{F}_A^{N }(q^2)$ the axial one. 
Since the interactions considered take place inside the SN core with a temperature $k_B T \ll m_N^*$ (where $k_B$ the Boltzmann constant that we set to $k_B=1$ from now on), for the scenario that we study $|q^2| \lesssim 0.15 \; \rm GeV^2$ and we can approximate $\tilde{F}_1^{N }(q^2) \sim \tilde{F}_1^{N }(0)$ and $\tilde{F}_A^{N }(q^2)\sim \tilde{F}_A^{N }(0)$. See Fig. 4.9 of~\cite{leitner}, where the dependence of the form factors on $|q^2|$ is shown, to understand better this approach. This is equivalent to say that at these energies neutrinos do not see the quark structure but the nucleon as a whole. Therefore, for our calculations we take $\tilde{F}_1^{n }(0) \sim -1/2$, $\tilde{F}_A^{n }(0) \sim -1.47/2$, $\tilde{F}_1^{p }(0) \sim 0.075/2$ and $\tilde{F}_A^{p }(0) \sim -1.47/2$~\cite{leitner}.

\section{Supernovae constraints on neutrino interactions}
\label{sec:SN}

Neutrinos play a key role in the first minutes after the collapse of a massive star. After these stars explode as SN, neutrinos are copiously produced and trapped in a core of radius $\sim 10$~km during the first $\sim 10$~s after the explosion \cite{Burrows:1986me, Janka2017}. At this time, the proto-NS core reaches densities several times that of nuclear saturation density, $\rho_B \sim  (2-3)  \rho_0$, and temperatures $T \lesssim 50$ MeV \cite{Pons_1999}. These neutrinos (and antineutrinos)  arise mostly from the electron capture on protons and flavour equipartition, i.e. electron $(\nu_e)$, muon and tau ($\nu_\tau,\nu_\mu$), collectively  $\nu_X$. Matter in the core is constituted of free nucleons, mostly neutrons, with which hot (thermalized) neutrinos interact many times before decoupling from matter inside the star. In this way, it is reasonable to think that new physics in the neutrino sector may alter SN dynamics in several ways. In particular, light mediators that couple to both neutrinos and nucleons could affect the diffusion time of neutrinos and their energy spectrum, which were first known from the measurements of the SN 1987A burst \cite{Bionta:1987qt, Hirata:1987hu, Alekseev:1987ej}.

New neutrino interactions can also modify the equation of state (EoS) of the proto-NS core \cite{Farzan:2018gtr}. This is particularly relevant if the new couplings between neutrinos and nucleons violate lepton number, as these processes can remove $\nu_e$ and convert them into $\bar{\nu}_e$, $\nu_\mu$, $\bar{\nu}_\mu$, $\nu_\tau$ and $\bar{\nu}_\tau$. It has been stated before \cite{fuller, Kolb:1981, rampp2002} that this type of interaction would imply sizeable differences in physical conditions, in particular in the entropy, lepton fraction and temperature, but these effects do not affect the explosion, as it is explained in Ref.~\cite{rampp2002}. In order to fully assess the effects of changes in the EoS, one must resort to numerical simulations, which is beyond the scope of this work. We therefore do not implement these bounds, but we show their potential reach.

The effects described above rely on the determination of the neutrino mean free path. As we argue in this section, this is a density and temperature dependent quantity, a fact that cannot be ignored in order to derive reliable bounds on new physics. In the following, we explain how this has been implemented in our calculation.

 \subsection{Mean free path in a nuclear medium}
\label{sec:mfp}
 
The mean free path in a dense and hot medium must be obtained from a self-consistent treatment, taking into account the distribution functions of all the particles of the interaction in the incoming and outgoing kinematic phase space at finite density and temperature.

The standard differential cross-section for the scattering between (anti)neutrinos, ($\bar{\nu}$) $\nu$, and nucleons, $N$,  can be written as
\begin{equation}
    d\sigma=\frac{|\mathcal{\overline{M}}|^2}{4\sqrt{(pk)^2-m_N^{*2} m_{\nu}^2}}d\Phi(p,p',k,k')(1-f_N(E'_N))(1-f_\nu(E'_\nu)),
    \label{eq:dsigma}
\end{equation}
where the phase space volume element is
\begin{equation}
    d\Phi(p,p',k,k')=(2\pi)^4\delta^{(4)}(p+k-p'-k')\frac{d^3\vec{p'}}{(2\pi)^32E'_N}\frac{d^3\vec{k'}}{(2\pi)^32E'_\nu}\ .
\end{equation}
The Fermi-Dirac distribution functions for the $i$th-type particle (protons and neutrons), $f_i(E_i)={1}/{1+e^{({E_i-\mueff_i})/{T}}}$, are written in terms of the effective mass and chemical potential $m^{\star}_i$ and $\mueff_i$, which differ from the naked values in vacuum by the presence of meson fields \cite{glendenning}. Here, $E_i$ is the nucleon energy and $T$ the temperature of the medium. The distribution function for the neutrino, $f_\nu(E_\nu)={1}/{1+e^{({E_\nu-\mu_\nu})/{T}}}$, depends on its Majorana or Dirac nature.

Note that the contribution from the scattering of neutrinos and antineutrinos off electrons is sub-leading \cite{Janka2017} and is therefore not included in our analysis.

Given the high values of the density and temperature that can be reached in a SN core, matter effects cannot be ignored when computing the neutrino mean free path. Vacuum expressions do not capture the rich physics of the many-body effects, which can increase the mean free path by more than two orders of magnitude~\cite{reddy, horow, cermeno1, cermeno2}. In our calculation, we have included the Fermi-Dirac distribution functions, which partially restrict the outgoing phase space, and the effective values for the nucleon mass (being $m_{N}^{\star} <m_N$) and chemical potential, thereby improving the results of previous works \cite{Farzan:2018gtr, Suliga:2020jfa}.

The neutrino mean free path can be expressed as the inverse of the cross section per unit volume, $\lambda=V/\sigma$. The differential cross section in \cref{eq:dsigma} has to be weighted with the phase space volume for the incoming nucleon sector, $2\frac{d^3\vec{p}}{(2\pi)^3}f_N(E_N)$. Neglecting the neutrino mass, $m_\nu \sim 0$, and performing a partial integration (see e.g., Ref.~\cite{cermeno1} for more details), we obtain
\begin{equation}
    \frac{d\sigma}{V}=\int
    \frac{|\mathcal{\overline{M}}|^2}{8(2\pi)^4|\vec{k}|\sqrt{(pk)^2}}\, |\vec{p}|\,\mathcal{F}(E_N, E'_N, E'_\nu)\,\delta (\cos\theta-\cos\theta^0)\, d\phi_{13}\, d|\vec{q}|\, dq_0\,  d|\vec{p}|\, d\cos\theta,
    \label{eq:dsigma2}
\end{equation}
where $\mathcal{F}$ is the Fermi blocking term 
\begin{equation}
    \mathcal{F}(E_N, E'_N, E'_\nu)=f_N(E_N)(1-f_N(E'_N))(1-f_\nu(E'_\nu)).
\end{equation}
We denote $\theta$ as the polar angle between $\vec{q}$ and $\vec{p}$, and $\theta_{ij}$ and $\phi_{ij}$ as the polar and azimuthal angles between the momentum of particle $i$ and particle $j$ (we take $i=1,\,2$ for the incoming neutrino and nucleon, respectively, and $i=3,\,4$ for the outgoing ones).

The angle $\theta^0$ is fixed by energy conservation, 
\begin{equation}
    \cos\theta^0=\frac{q_0^2+2q_0E_N-|\vec{q}|^2}{2|\vec{q}||\vec{p}|},
\end{equation}
and the flux term can be expressed as
\begin{equation}
    \sqrt{(pk)^2}=E_N E_\nu- |\vec{p}||\vec{k}|\cos\theta_{12}.
    \label{eq:flux}
\end{equation}
The polar angle, $\theta_{12}$, between the incoming neutrino and nucleon is obtained by solving
\begin{equation}
    A  \; \sin \theta_{12}+ B \; \cos \theta_{12}+ C=0,
\end{equation}
with $A=|\vec{p}||\vec{k'}|\sin\phi_{13}\sin\theta_{13}$, $B=|\vec{p}|(|\vec{k'}|\cos\theta_{13}-|\vec{k}|)$ and $C=|\vec{p}||\vec{q}|\cos\theta$.
Note that $|\vec{k}|=E_\nu$, $|\vec{k'}|=E'_\nu=E_\nu-q_0$, and $E'_N=E_N+q_0$. 
After doing this, \cref{eq:dsigma2} can be rewritten as
\begin{equation}
    \frac{\sigma}{V}=\frac{1}{8(2\pi)^4}\int_{- \infty}^{E_\nu} d q_0 \int_{|q_0|}^{2E_\nu-q_0}d|\vec{q}| \int_0^{2\pi} d\phi_{13} \int_{|\vec{p}_-|}^{\infty}d|\vec{p}||\vec{p}|\mathcal{F}(q_0, |\vec{p}|)\frac{|\mathcal{\overline{M}}|^2   }{E_\nu(E_N E_\nu- |\vec{p}|E_\nu\cos\theta_{12})},
    \label{eq:mfp}
\end{equation}
with $|\vec{p}_{-}|=\sqrt{E_{N-}^2-m_N^{*2}}$ and $E_{N-}=\frac{-q_0}{2}+\frac{|\vec{q}|}{2} \sqrt{1-\frac{4m_N^{*2}}{q^2}}$. Note that the limits of the integral over $|\vec{p}|$ comes from imposing $|\cos\theta^0|\leq 1$.

The density profile of the proto-NS and the temperature are radially dependent functions. In the centre it is well above the nuclear saturation density and decreases towards the outer regions. As a consequence, the neutrino energy is also a function of the radius, and it changes within the hot and dense medium. 
A complete treatment should also include the time-dependence of these quantities, but this is only possible with a dedicated numerically simulation, which is beyond the scope of this  paper. 
As a more modest approach, we use the results from Ref.~\cite{Fischer:2011cy}, where the radial and time dependence of the proto-NS temperature, density, etc, were computed for an $18\,M_\odot$ progenitor, considering the TM1 model \cite{Sugahara:1993wz, Shen:1998gq} to describe the physics of nuclear matter. As we are most interested in processes during the Kevin-Helmholtz phase, we have used results at times of $1$ and $5$ seconds post-bounce.

\begin{center} 
\renewcommand{\arraystretch}{1.2}
\begin{table}[t!]
\begin{tabular}{l|c|c|c|c||c|c|c|c||c|c}
\multicolumn{4}{l}{Dirac Neutrinos  \rule{0ex}{2.6ex}}  \\ 
\hline
\hline
$t \sim 1$ s& $R$ (km) &$T$ (MeV) & $n_B (\rm fm^{-3})$ & $Y_e$&$\mueff_n$ (MeV) & $\mueff_p$ (MeV) & $\mu_\nu$ (MeV)& $m_{N}^{\star}$ (MeV) & $\lambda_{\nu_e}^{\rm SM}$ (m) & $\lambda_{\bar{\nu}_e}^{\rm SM}$ (m) \rule{0ex}{2.6ex}
\\ \hline
$k=1$& 5.0&
15 & 0.5 & 0.3
& 496.6 & 405.4 & 114.6 & 249.6 & 0.42 & 0.30
\\ \hline
$k=2$& 7.5&
20 & 0.3 & 0.28
& 530.0 & 458.3 & 102.7 & 384.9 & 0.24 & 0.22
\\ \hline
$k=3$& 10.0&
28& 0.15 & 0.25
& 656.5 & 601.9 & 79.9 & 599.4 & 0.17 & 0.20
\\ \hline
$k=4$& 15.0&
33& 0.06 & 0.2
& 779.8 & 723.0 & 29.0 & 786.0 & 0.42 & 0.49
\\ \hline
$k=5$& 17.5&
18& 0.03 & 0.1
& 858.7 & 813.1 & 14.4 & 857.0 & 2.7 & 2.9
\\ \hline
$k=6$& 20.0&
7& 0.008 & 0.05
& 917.2 & 893.9 & 12.5 & 915.9 & 35 & 36
\\ 
\hline
\hline
$t \sim 5$ s\rule{0ex}{2.6ex}& &&&&&&&& 
\\ \hline
$k=1$& 5.0&
25 & 0.5 & 0.25
& 504.3 & 389.4 & 41.8 & 254.6 & 0.30 & 0.38
\\ \hline
$k=2$& 7.5&
28 & 0.4 & 0.23
& 509.4 & 402.4 & 36.1 & 309.1 & 0.40 & 0.50
\\ \hline
$k=3$& 10.0&
32& 0.3 & 0.2
& 537.7 & 440.2 & 24.4 & 394.4 & 0.21 & 0.28
\\ \hline
$k=4$& 12.5&
25& 0.15 & 0.13
& 664.8 & 590.6 & 14.2 & 599.1 & 0.51 & 0.63
\\ \hline
$k=5$& 15.0&
10& 0.05 & 0.035
& 831.9 & 787.2 & 0 & 805.3 & 12 & 13
\\ 
\hline
\hline
\end{tabular}
    \caption{Values of neutron effective chemical potential, $\mueff_n$, proton effective chemical potential, $\mueff_p$, neutrino chemical potential, $\mu_\nu$, and nucleon effective mass, $m_{N}^{\star}$, for the spherical shells (labeled by the index $k$ and defined by an outer radius $R$) that we consider at the two time snapshots of $1$ s and $5$ s, with a baryonic density, $n_B$, temperature, $T$ and electron fraction, $Y_e$, in our analysis for Dirac neutrinos. In the last two columns the resulting neutrino mean free path in the SM, $\lambda_{\nu_e}^{\rm SM}$, and antineutrino mean free path, $\lambda_{\bar{\nu}_e}^{\rm SM}$, are also shown. Temperatures, densities and electron fraction are taken from Ref.\,\cite{Fischer:2011cy}.
}
\label{tab:mfp_2}

\begin{tabular}{l|c|c|c||c|c|c||c|c}
\multicolumn{4}{l}{Majorana Neutrinos  \rule{0ex}{3.6ex}}  \\ 
\hline
\hline
$t \sim 1$ s& $R$ (km)& $T$ (MeV) & $n_B (\rm fm^{-3})$ &$\mueff_n$ (MeV) & $\mueff_p$ (MeV) &  $m_{N}^{\star}$ (MeV) &
\begin{minipage}{1.9cm}
    $\lambda^{\rm SM}$ (m) \rule{0ex}{2.6ex}\\ 
    $\left(E_\nu = \pi T\right) \rule{0ex}{2.6ex}$\\
\end{minipage}
& 
\begin{minipage}{2.4cm}$
    \lambda^{\rm SM} $ (m) \rule{0ex}{2.6ex}\\ 
    $\left(E_\nu = \mu_\nu^D+ \pi T\right)\rule{0ex}{2.6ex}$ 
\end{minipage}
\\ \hline
$k=1$& 5.0
& 15 & 0.5 
& 512.3 & 382.9 & 253.9 & 3.1 & 0.27
\\ \hline
$k=2$& 7.5
& 20 & 0.3 
& 544.4 & 440.6 & 389.7 & 1.3 & 0.21
\\ \hline
$k=3$& 10.0
& 28& 0.15 
& 538.0 & 431.4 & 383.0 & 0.51 & 0.17
\\ \hline
$k=4$& 15.0
& 33& 0.06 
& 781.9 & 713.3 & 786.1 & 0.69 & 0.46
\\ \hline
$k=5$& 17.5
& 18& 0.03 
& 859.2 & 808.2 & 857.0 & 4.1 & 2.7
\\ \hline
$k=6$& 20.0
& 7& 0.008 
& 917.4 & 889.9 & 915.9 & 76 & 35
\\ 
\hline
\hline
$t \sim 5$ s \rule{0ex}{2.6ex}& &&&&&&
\\ \hline
$k=1$& 5.0 &
25 & 0.5 
& 509.9 & 381.1 & 256.2 & 0.71 & 0.34
\\ \hline
$k=2$& 7.5 &
28 & 0.4 
& 514.3 & 395.4 & 310.8 & 0.52 & 0.29
\\ \hline
$k=3$& 10.0 &
32& 0.3 
& 541.0 & 435.5 & 395.4 & 0.36 & 0.25
\\ \hline
$k=4$& 12.5 &
25& 0.15 
& 665.9 & 587.1 & 599.3 & 0.77 & 0.58
\\ \hline
$k=5$& 15.0 &
10& 0.05 
& 831.9 & 787.0 & 805.3 & 4.1 & 4.1
\\ 
\hline
\hline
\end{tabular}
    \caption{The same as \cref{tab:mfp_2} but for Majorana neutrinos. The neutrino mean free path is calculated for two choices of the neutrino energy, $E_\nu =  \pi T$ and $E_\nu = \mu_\nu^D+ \pi T$. 
}
\label{tab:mfp_2_maj}
\end{table}
\end{center}

Since the density decreases slowly within the core, we have considered spherical shells where temperature and density are taken to be constant, with values extracted from Fig.~7 of Ref.~\cite{Fischer:2011cy} as detailed in \cref{tab:mfp_2,tab:mfp_2_maj}. We consider these shells from the centre of the star to the neutrino sphere radius, $R_\nu$,  where neutrinos decouple from matter. The neutrino sphere radius is a function of the neutrino energy, $E_\nu$, flavour and thermodynamical state of the matter \cite{Janka2017, Fischer:2011cy}, and therefore depends on the time after bounce. Knowing the antineutrino spectrum from SN 1987A, which suggests that neutrinos decoupled with energies around a dozen MeV or, accordingly, temperatures of a few MeV (see Fig.~1 in Ref.~\cite{Janka2017}),
the neutrino sphere can be determined from the following condition on the optical depth, $\tau_\nu$,
\begin{equation}
    \tau_\nu(E_\nu)=\int_{R_\nu(E_\nu)}^\infty dr \frac{1}{\lambda(r, E_\nu)}=\frac23\ .
    \label{eq:tau}
\end{equation}
Neutrinos with purely SM interactions (i.e., the electroweak force) already fulfil the opacity constraint and they remain diffusive while streaming out. Adding a new physics contribution to neutrino interactions increases, in general, the probability of scattering, and therefore the condition of Eq.~\eqref{eq:tau} is satisfied. The addition of a new vector mediator, however, must be considered more carefully, as the destructive interference with the $Z$ boson can dominate in certain regions of the parameter space. We have checked explicitly that neutrinos remain diffusive in all the parameter space.

Regarding the neutrino energy, we consider Dirac neutrinos to be in thermal equilibrium with an energy $E_\nu=\mu_\nu+\pi T$ \cite{Pons_1999}, where $\mu_\nu$ is the neutrino chemical potential. Majorana neutrinos do not reach thermal equilibrium (and do not have a chemical potential associated to them). Their energy should be determined numerically, but at the time of writing this article, we are not aware of any hydrodynamic simulation of SN with Majorana neutrinos. Therefore, we consider two limiting cases for the Majorana neutrino energy, namely $E_\nu = \pi T$ (a relation that has been used, e.g., in Ref.~\cite{Kolb:1981}) and $E_\nu=\mu_\nu^D+\pi T$, where $\mu_\nu^D$ is the chemical potential for Dirac neutrinos.

In order to determine the effective nucleon mass and chemical potentials in each shell, we consider the TM1 model (also used in Ref.~\cite{Fischer:2011cy} to obtain the radial temperature and density profiles that we use in \cref{tab:mfp_2,tab:mfp_2_maj}). We use a relativistic mean field (RMF) approach where baryons are considered as Dirac quasiparticles moving in classical meson fields and the field operators, $\phi$, are replaced by their expectation values, $\langle \phi \rangle$. The TM1 \cite{Sugahara:1993wz, Shen:1998gq} model, widely used in current numerical simulations, is a representative example where the set of parameters used can smoothly connect low and high density regions in the dynamical stellar description. The presence of an effective nucleon mass and effective chemical potential is due to the non vanishing values of the Lorentz scalar meson, $\langle\sigma\rangle$, Lorentz vector, $\langle\omega_\mu\rangle$, and vector-isovector, $\langle \vec{\rho}_\mu\rangle$, meson fields. The self-consistent solution in the RMF approximation of the meaningful components of these field values  $(\sigma,\omega_0,\rho_{03})$ was obtained in Refs.\,\cite{Sugahara:1993wz, Shen:1998gq}. In this way, an effective nucleon mass,  $m_N^{*}=m_N-g_{\sigma N} \langle \sigma \rangle$, and effective chemical potentials, $\mu_{i}^{*}=\mu_{i}-g_{\omega N} \langle\omega \rangle-g_{\rho N} t_{3 i} \langle\rho\rangle$, are obtained for each thermodynamical state. The quantities $g_{\sigma N}$, $g_{\omega N}$, and $g_{\rho N}$ are dimensionless constants that couple nucleons to the $\sigma$, $\omega$, and $\rho$ mesons, respectively. $t_{3 i}$ is the third component of the isospin  of
the proton or the neutron, $i=p,n$. Relativistic baryonic energies are
defined as $E_{i}^{*}(k)=\sqrt{k^{2}+m_{N}^{\star 2}}$. This parameter set includes self-interaction terms from scalar, vector and vector-isovector mesons in non isospin symmetric nuclear matter at finite temperature. TM1 interaction terms are constrained by the nuclear masses, radii, neutron skins and their excitations. When applied to the derived proto-NS, the  mass-radius diagram allows to fulfil the subsequent two solar mass constraint from recent observations of older objects.

By imposing baryonic charge conservation, fixed lepton fraction and charge neutrality, the RMF equation for the fields, $\sigma$, $\omega$, and $\rho$ are solved to obtain effective nucleon masses and chemical potentials at finite temperature \cite{Sugahara:1993wz}. Note that at finite temperature a non-vanishing positron fraction is available and a net lepton number (including the neutrino sector) arises. The resulting values of $m_{N}^{\star}$ and chemical potentials are shown in \cref{tab:mfp_2,tab:mfp_2_maj}. For reference, we also indicate the resulting neutrino mean free path in the SM, $\lambda^{\rm SM}$. For the Majorana case, we show the two values that correspond to the limiting cases for the energy window we consider, $E_\nu = \pi T$ and $E_\nu=\mu_\nu^D+\pi T$.

Finally, using these values of the effective masses and chemical potentials in \cref{eq:mfp}, we compute the neutrino and antineutrino mean free paths for each spherical shell, $\lambda_k$, to determine the diffusion time in each of them. We add these to estimate the total duration of the neutrino (antineutrino) emission as
\begin{equation}
    c\Delta t^{\nu (\bar{\nu})} = \sum_{k=1}^n \frac{R_k^2-R_{k-1}^2}{\lambda_k^{\nu (\bar{\nu})}},  
    \label{eq:deltat}
\end{equation}
where $R_k$ is the outer radius of the $k$-th spherical shell, and $R_0 = 0$. Since the neutrinos detected in the signal from SN1987A were electron antineutrinos we compare the total duration of the antineutrino emission with the observed $\Delta t\sim10$~s.

\subsection{Effects on SN dynamics }
\label{sec:diff}

During the core collapse, protons and electrons combine into neutrons, with each of the $\mathcal{O}(10^{57})$ reactions producing an associated electron neutrino but also other flavours (muon and tau) and reactions come into play including electron-positron annihilation, or nucleon bremstrahlung \cite{Hannestad:1997gc}.   Many of these neutrinos and antineutrinos escape the developing neutron star, carrying away the largest portion of the gravitational energy lost in the collapse. This neutrino burst was detected by the experiments Kamiokande II, IMB, and Baksan in coincidence with the SN 1987A event \cite{Bionta:1987qt, Hirata:1987hu, Alekseev:1987ej} and their measurement allows us to place constraints on models of new physics which would affect the propagation of the neutrino burst through the proto-NS. In particular, the diffusion time scale $\Delta t$ of \cref{eq:deltat} can be altered by new interactions between neutrinos and nucleons. One must impose that this quantity is of the order of  the observed duration of the burst, $\Delta t \lesssim 10 \; \rm s$.

When we apply \cref{eq:deltat} using pure vector-axial SM interactions, we obtain $\Delta t^{\rm SM} = 2.6$~s for Dirac neutrinos and $\Delta t^{\rm SM} =2.4$~s for antineutrinos. For Majorana neutrinos, we obtain $\Delta t^{\rm SM} =1.1$~s  ($\Delta t^{\rm SM} =2.7$~s) for $E_\nu = \pi T$ ($E_\nu=\mu_\nu^D+ \pi T$). For consistency, these numbers have been obtained using the configuration for $t \sim 1$~s of \cref{tab:mfp_2,tab:mfp_2_maj} \footnote{If we had used the configuration for $t \sim 5$~s we would have obtained $\Delta t^{\rm SM} = 1.5$~s ($1.3$~s) for Dirac neutrinos (antineutrinos) and $\Delta t^{\rm SM} =1.6$~s ($1.0$~s) for Majorana neutrinos with $E_\nu^M = \pi T$ ($E_\nu^M=\mu_\nu^D+ \pi T$).}.

The mean free path for Majorana neutrinos is extremely sensitive to the incoming neutrino energy. If the same energy as for Dirac neutrinos is used, Majorana and Dirac neutrinos yield a very similar $\lambda_k^{SM}$. The main difference between these two scenarios comes from the energy distribution of neutrinos in the star, and if $E_\nu = \pi T$ is used, the neutrino mean free path for Majorana neutrinos increases considerably (by up to a factor $\sim 4$, depending on the SN shell). To obtain a more precise answer, neutrino energy should be determined by numerical simulations, especially for Majorana neutrinos, which do not reach thermal equilibrium.

We should point out that we have not considered correlations that typically enhance the mean free path, see for example Ref.~\cite{horow}, where random phase approximation correlations indicate that in-medium instabilities cause an increased mean free path. In addition, the inclusion of weak magnetism effects can alter these values by approximately $20\%$. A fully consistent treatment of the dispersion suffered by neutrinos would require the use of Boltzmann transport and energy conservation at each interaction step. Since this is not feasible in our calculation, we rely on semi-analytical estimates to include the size of corrections expected when new physics is involved. This, together with the choice of a pure axial-vector current for the SM interaction description (realistic simulations consider extra tensorial terms \cite{horow}) might explain why our results for the SM diffusion time is slightly shorter than $10$~s.

After new physics contributions are included, we need to ensure that the total neutrino diffusion time, $\Delta t$, remains consistent with the duration of the observed SN1987A neutrino burst, which leads to the condition
\begin{equation}
    \Delta t \lesssim 10 \; \rm s, 
    \label{eq:diff1}
\end{equation}
setting an upper limit on the strength of neutrino interactions with quarks. Since what experiments have observed is the burst of antineutrinos through the measurement of positrons coming from $\bar{\nu}_e n \rightarrow p e^+$, we apply this constraint to the antineutrino diffusion times. Lacking a full simulation analysis, the systematic error of this numerical approach is difficult to estimate. Since the SM times that we obtain are slightly shorter than the $10$~s observed, we also indicate the region in which the new physics contribution becomes as important as the SM one, which can be estimated as follows,
\begin{equation}
    \Delta t  \lesssim 2\, \Delta t^{\rm SM}\ .
    \label{eq:diff2}
\end{equation} 
For consistency, the bound given by \cref{eq:diff1} is computed using the $t\sim 5$~s configuration of \cref{tab:mfp_2,tab:mfp_2_maj}, whereas for the one of \cref{eq:diff2} we must use the $t\sim 1$~s configuration, so that the stellar properties correspond to those at half the amount of time that neutrinos typically spend in the supernova.

If the interaction between neutrinos and quarks via a scalar mediator is LNC, right handed neutrinos with masses $m_{\nu_R} \lesssim 1$~eV can be produced in the scattering of left handed neutrinos off SM fermions. Once $\nu_R$ are produced, they free stream out of the star since they do not interact via SM interactions, leading to a suppressed $\nu_L$ flux and a much shorted cooling time. This can be prevented in two ways, either $\nu_L$ do not interact with SM fermions via a new physics interaction while they are trapped in the proto-NS, i.e.,
\begin{equation}
    \sum_{k=1}^n \frac{c\Delta t_k}{\lambda_k^{\mathrm{NP}}} \lesssim 1,
    \label{eq:eos}
\end{equation}
with 
\begin{equation}
    c\Delta t_k \sim c\Delta t_k^{\rm SM}=\frac{R_k^2-R_{k-1}^2}{\lambda_k^{\rm SM}},
\label{eq:deltak}    
\end{equation}
where $\lambda_k^{\rm NP}$ the mean free path computed only with new physics interactions. Alternatively, if the new physics scattering rate is sufficiently large, the $\nu_R$ do not free stream out of the core. Instead, they scatter back into $\nu_L$, and the overall effect is that the neutrinos are constantly switching back and forth between $\nu_L$ and $\nu_R$ as they diffuse out of the SN. In line with the analysis of Ref.~\cite{Farzan:2018gtr}, we constrain this effect by requiring that at least $100$ new physics interactions take place over the path the $\nu_R$ would take when free streaming out of the SN, this is,
\begin{equation}
    \sum_{k=1}^{n} \frac{\Delta R_{k}}{\lambda_k^{\mathrm{NP}}} \gtrsim 100\,,
    \label{eq:nu_R_trapping}
\end{equation}
where $\Delta R_{k}=R_{k}-R_{k-1}$. Note that, in this case, neutrinos spend on average half of their time inside the star as $\nu_R$ and not feeling electroweak interactions. Thus, the diffusion time is calculated as 
\begin{equation}
    c \Delta t=\frac{1}{2}  \sum_{k=1}^n \left(R_k^2-R_{k-1}^2\right) 
    \left[\frac{2}{\lambda_k^{NP}} + \frac{1}{\lambda_k^{SM}} \right]\, .
\end{equation}
The condition in \cref{eq:nu_R_trapping} must be understood as a qualitative statement, limiting the relative size of the new physics and SM contributions.\footnote{Alternatively, a similar condition could be implemented on the radius of the right-handed neutrinosphere, constraining by how much it can exceed the radius of the left-handed neutrinosphere.}

Finally, as we have mentioned above, if new neutrino physics happens via a LNV interaction, quantities as the entropy, lepton fraction and temperature can be greatly affected. Qualitatively speaking, one can estimate when neutrino interactions might have an impact on the EoS of SN matter by requiring that they scatter at least once as they stream out of the star. This minimal condition can be imposed as in the LNC case, requiring that \cref{eq:eos} is fulfilled.

Changes in the EoS of SN matter would only indirectly affect the neutrino signal: either through a change of detected flavours or time correlation on Earth. Currently, preliminary microscopic simulations have been performed using one neutrino species allowing internal deleptonization with $\sigma/\sigma_{SM}\lesssim 10^{-3}$ for the reaction $\nu_e+N \rightarrow {\bar \nu_e}+N$  \cite{Rampp:2002kn}. It should be noted that this work, based on one single flavour LNV reactions, employ one-dimensional codes where luminosities are monitored for moderate times after the trigger of the stellar collapse. Because of their simplified microphysics input, they do not display all the physics features relevant to the dynamics and outcome of the observable signals at Earth, such as convection \cite{Mezzacappa:1997gua} and pre-collapse seed perturbations \cite{Burrows_2018, Burrows_2021}. The results of Ref.~\cite{Rampp:2002kn} suggest some degree of insensitivity of the overall dynamical evolution of the models to rather dramatic modifications of the microphysics, but to obtain a reliable result, it is necessary to extend the simulation to larger times.

More recent dedicated studies have performed detailed numerical simulations that include other exotic species, such as sterile neutrinos with ${\cal O}(50)$~MeV masses that decay into SM neutrinos. It was found that these could have a significant impact on the dynamics of the core collapse.  In fact, for some parameter choices, they can lead to too energetic explosions thus indicating that with the help of the astrophysics data it is possible to further constrain the parameter space of sterile neutrinos~\cite{Rembiasz:2018lok}. Given the proto-NS star conditions described in our contribution we anticipate that the early deleptonization could be very different from that predicted using SM neutrino transport. Further work on the computational side is needed in order to obtain the full picture regarding how the final spectra would look like as measured in a hypothetical neutrino detection on Earth.

Because of the arguments above, and in line with previous literature \cite{Farzan:2018gtr}, we only indicate when the condition given by \cref{eq:eos} is fulfilled for an LNV model in \cref{fig:LNV_constraints}, but we do not consider it a solid constraint yet. In order to obtain robust results, a full hydrodynamic simulation would be needed that relates changes in the EoS with SN observables.

\vspace*{1.5ex}
Previous works \cite{Farzan:2018gtr, Suliga:2020jfa} have implemented the conditions of \cref{eq:diff1,eq:diff2,eq:eos,eq:deltak,eq:nu_R_trapping} to constrain new physics models, but ignoring the Pauli blocking for the outgoing nucleons and their effective masses and chemical potentials. In our work, we use the results of the previous section, where medium effects have being taken into account, for the calculation of the mean free path in \cref{eq:mfp} to derive a more consistent estimate.

\section{Discussion}
\label{sec:discussion}

In this section, we compute the neutrino mean free path for the new physics models of \cref{sec:models} and show the constraints on the corresponding parameter spaces.

First, in order to quantify the importance of medium effects on our results, we compare the values of the mean free path in the $k$-th shell of the star including medium effects ($\lambda_k$, obtained using \cref{eq:mfp}) with those found neglecting both the Pauli blocking of the outgoing states and the effect of the dense nuclear medium on the nucleon mass ($\lambda_k^{0}$). To compute $\lambda_k^{0}$ one must include the density of nucleons in each shell. To mimic vacuum calculations, where $\lambda^{vac} \sim (\sigma n_B)^{-1}$, we perform the integral over the phase space volume for the incoming nucleon sector, $n_B \sim 2{d^3\vec{p}}  f_N(E_N)/{(2\pi)^3}$, retaining the effective mass of the nucleon only in this factor for consistency with the baryonic density of each shell (note that the effective mass is related to the baryonic density through the RMF equations). Since we are keeping $f_N$, $\lambda_k^0$ is not truly a {\em vacuum} mean free path and it incorporates some in-medium effects.

\begin{figure}
    \centering
    \includegraphics[width = 1.\textwidth]{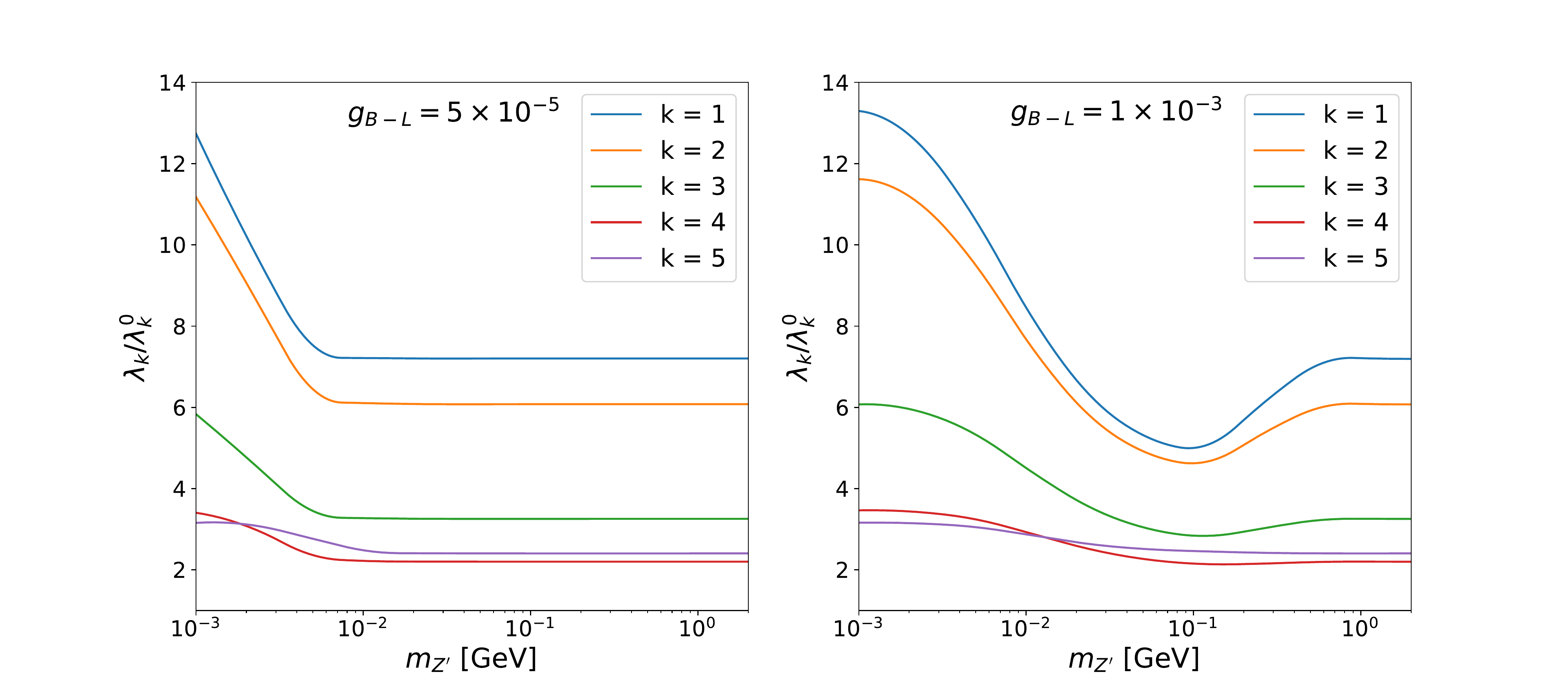}
    \includegraphics[width = 1.\textwidth]{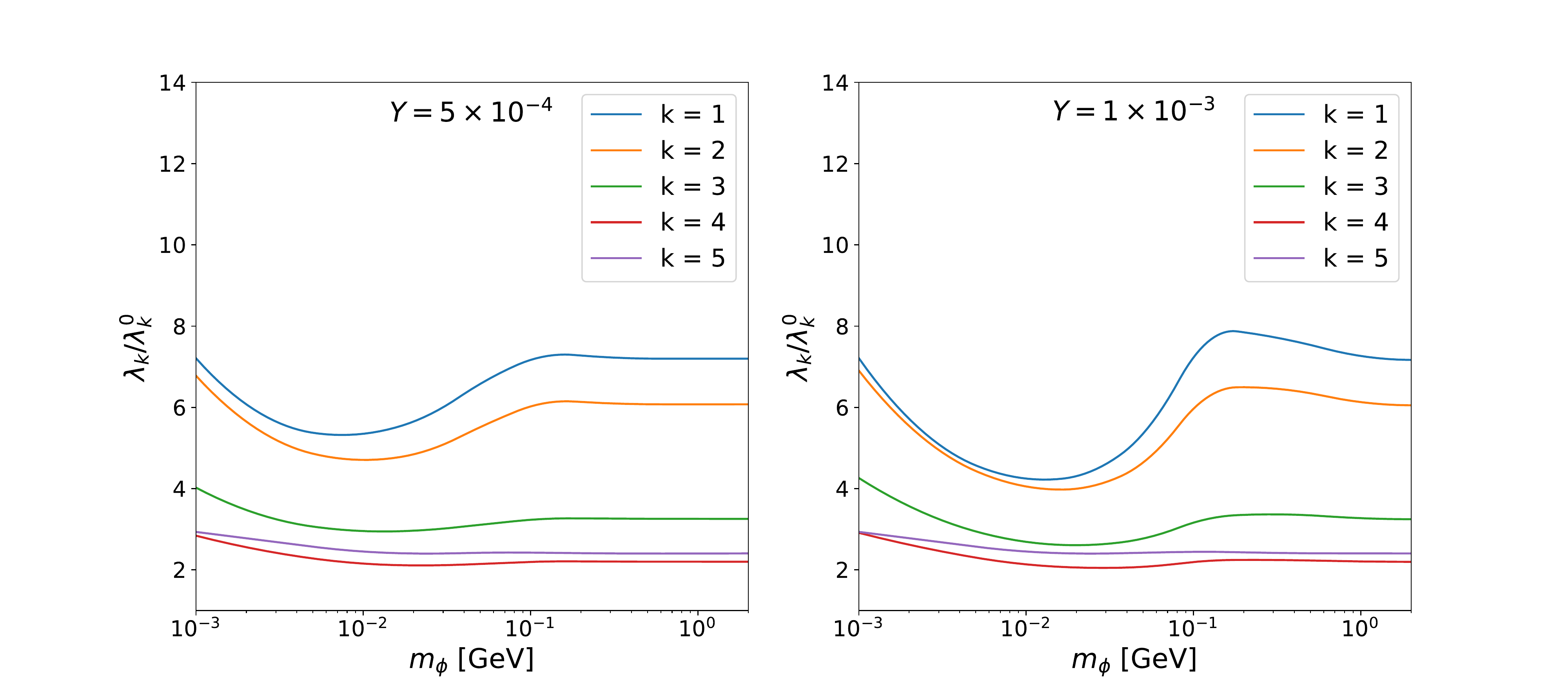}
    \vspace*{-3ex}
    \caption{Ratio of the values of the mean free path in the $k$-th shell of the star including medium effects ($\lambda_k$) with those found neglecting the Pauli blocking term and using the vacuum value for the nucleon masses ($\lambda_k^0$, see text for more details) for the $t \sim 5$ s snapshot. The upper plots correspond to \ubl model while the bottom ones are related to the LNC scalar mediated model. For each model we consider two different values for the coupling. The left plots correspond to couplings providing new physics interactions which do not distort the SM scenario for large masses while the right ones illustrate cases where new physics interactions affect the diffusion time.}
    \label{fig:medium}
\end{figure}

\begin{figure}[t!]
  \centering
  \includegraphics[width=.45\columnwidth]{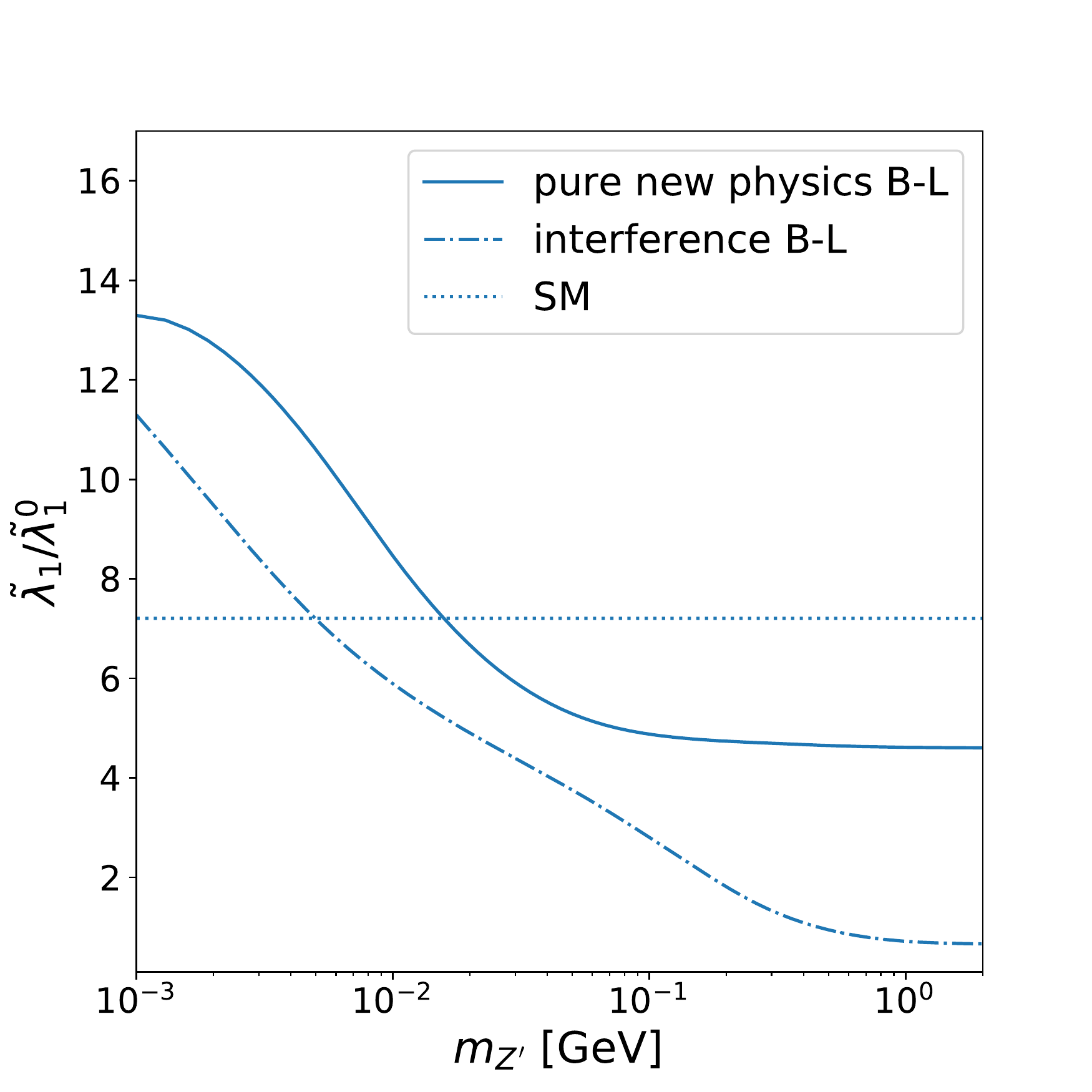} \hspace{-0.75cm}
  \includegraphics[width=.45\columnwidth]{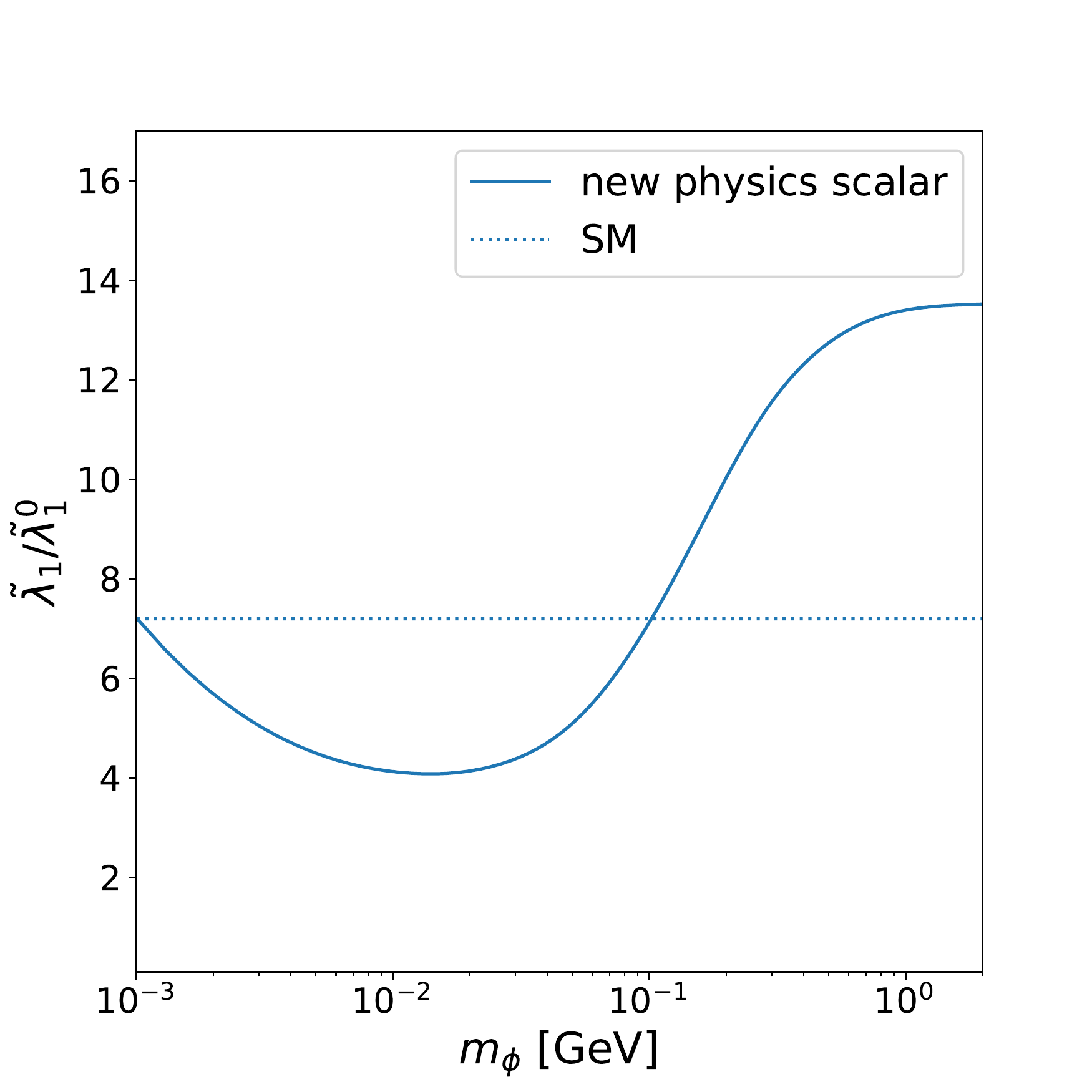}
  \caption{Ratio of the contributions of the new physics and SM terms to the mean free path in the 1st shell of the star including medium effects ($\tilde{\lambda}_1$) with those found neglecting the Pauli blocking term and using the vacuum value for the nucleon masses ($\tilde{\lambda}_1^0$) for a $t \sim 5$ s snapshot. Left plot for the \ubl model and right plot for the LNC scalar mediated one.}
   
 \label{fig:medium_npvssm}
    
\end{figure}
In \cref{fig:medium}, we plot the ratios $\lambda_k/\lambda_k^0$ as a function of the mediator mass for two representative choices of the coupling in both the $U(1)_{B-L}$ and LNC scalar models, and for the five radial shells at $t=5$~s described in \cref{tab:mfp_2}. We can observe that medium effects are more important in the inner shells (where the temperature and density are higher) and decrease as we move outwards in the SN. The neutrino mean free path can increase by approximately one order of magnitude with respect to the vacuum estimate in the central region of the SN, an increase which is more pronounced (up to a factor 14) with new physics contributions. In each plot, the lines become horizontal in the limit where the SM dominates over the new physics contributions, which happens at small values of the couplings and large mediator masses.

The non-trivial behaviour of $\lambda_k/\lambda_k^0$ when new physics contributions dominate is due to the dependence of the scattering amplitudes of \cref{eq:Msq_scalar,eq:MBL,eq:Mmixingnu,eq:Mmixingantinu} on the effective nucleon mass and mediator mass. The mediator mass enters through the denominator of the corresponding propagator (${1}/(q^2-m_\phi^2)$ or ${1}/(q^2-m_{Z'}^2)$), affecting the range of the $q^2$ values that enter the integration in \cref{eq:mfp}, which also multiply terms in the numerator with the effective nucleon mass. For heavy mediators ($m_{Z',\phi}^2\gg|q^2|$), this asymptotically reaches a flat behaviour. This can be seen more clearly in \cref{fig:medium_npvssm}, where we represent the ratio $\lambda_k/\lambda_k^0$, computed separately for each contribution (new physics, interference term, and SM) for the innermost shell (a quantity that we define as $\tilde\lambda_1/\tilde\lambda_1^0$). The non-trivial behaviour observed in \cref{fig:medium} appears when all the contributions are included in the calculation of the mean-free path.

\begin{figure}[!t]
    \centering
    \includegraphics[width=.8\textwidth]{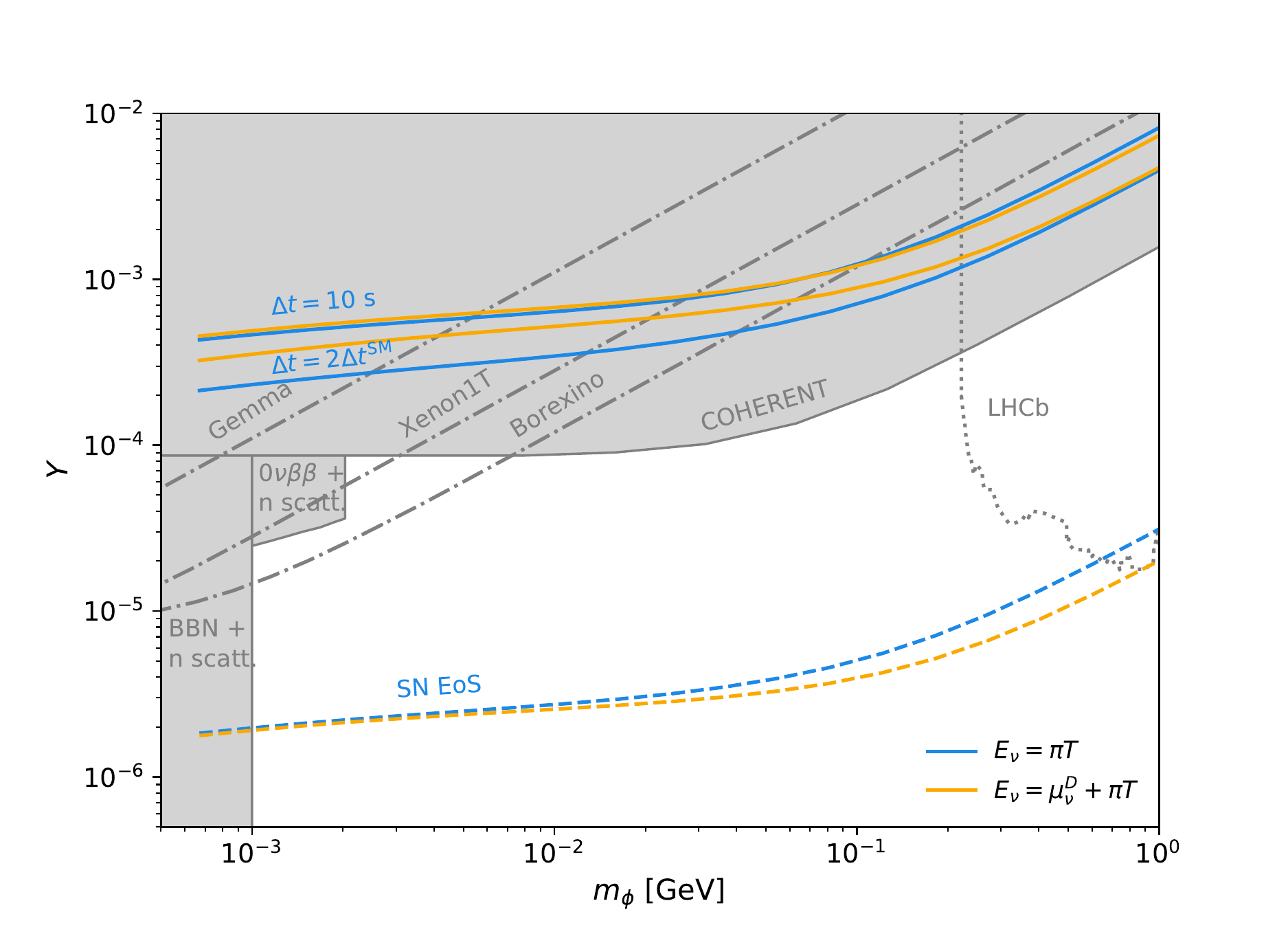}
    \vspace*{-2ex}
    \caption{Constraints on an LNV scalar mediator. Solid lines represent direct constraints on the coupling $Y=\sqrt{C_\nu C_N}$, dot-dashed lines represent those translated from constraints on $\sqrt{C_\nu C_e}$, and dotted lines are from constraints on $\sqrt{C_\mu C_q}$. These latter two classes of constraints can only be compared with the SN limits in a model-dependent way. Here they are plotted under the assumption of universal couplings to all SM fermions. For models with different couplings the lines can be rescaled, as described in the text. Constraints from supernova neutrinos are shown under two possible assumptions for the neutrino energy: $E_\nu=\pi T$ (blue), and $E_\nu = \mu_\nu^D + \pi T$ (orange). The two solid lines of each colour correspond to two possible criteria under which the neutrino diffusion time can be constrained: $\Delta t = 10$ s and $\Delta t = 2 \Delta t^{\rm{SM}}$, while the dashed lines represent the region above which the SN EoS is likely to be affected by the presence of LNV interactions.
    }
    \label{fig:LNV_constraints}

    \vspace*{-1.ex}
    \centering
    \includegraphics[width=.8\textwidth]{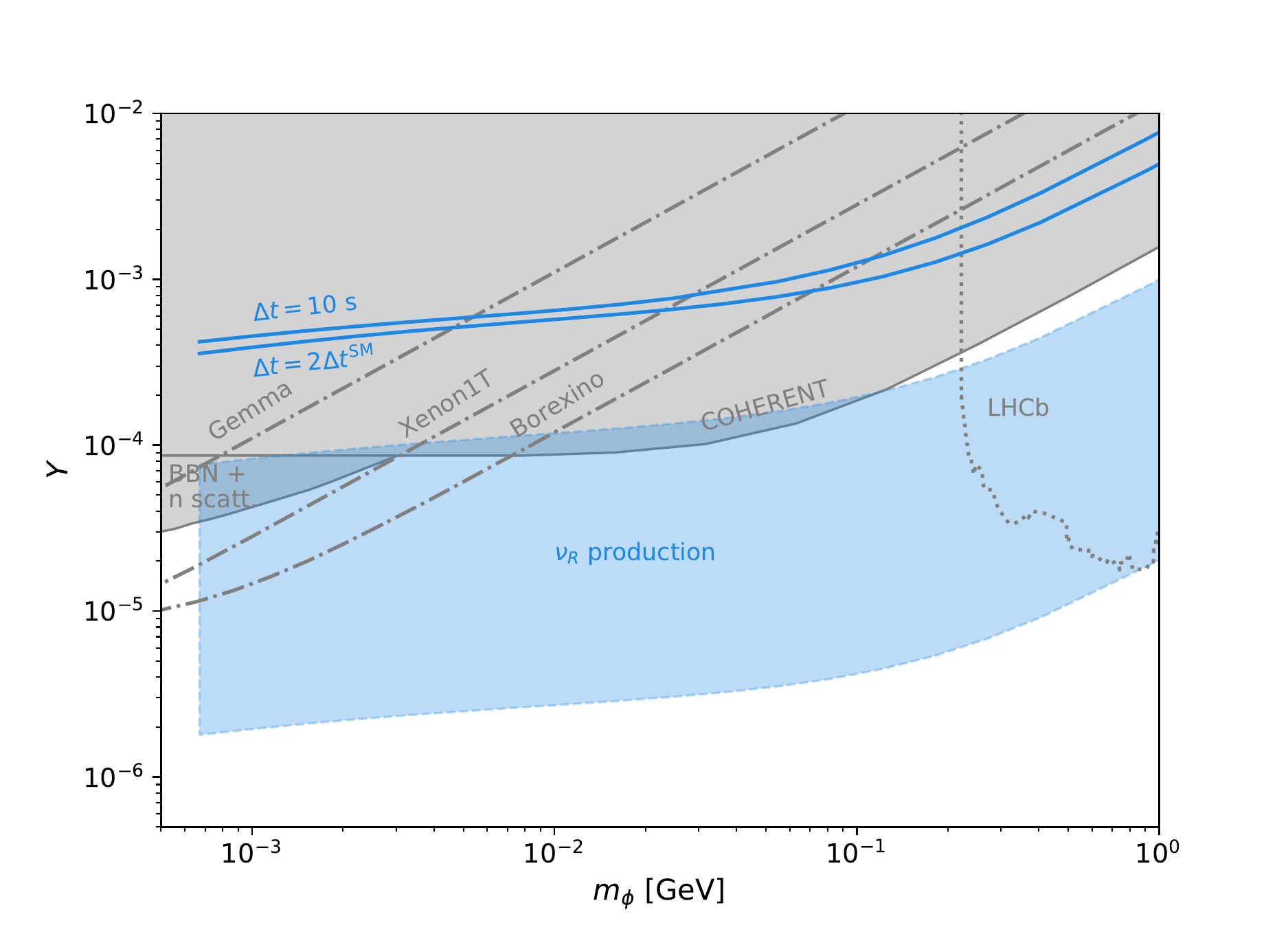}
    \vspace*{-2ex}
    \caption{The same as in \cref{fig:LNV_constraints}, but for an LNC scalar. The shaded blue area is ruled out because of right handed neutrino production.}
    \label{fig:LNC_constraints}
\end{figure}

\begin{figure}[!t]
    \centering
    \includegraphics[width=.8\textwidth]{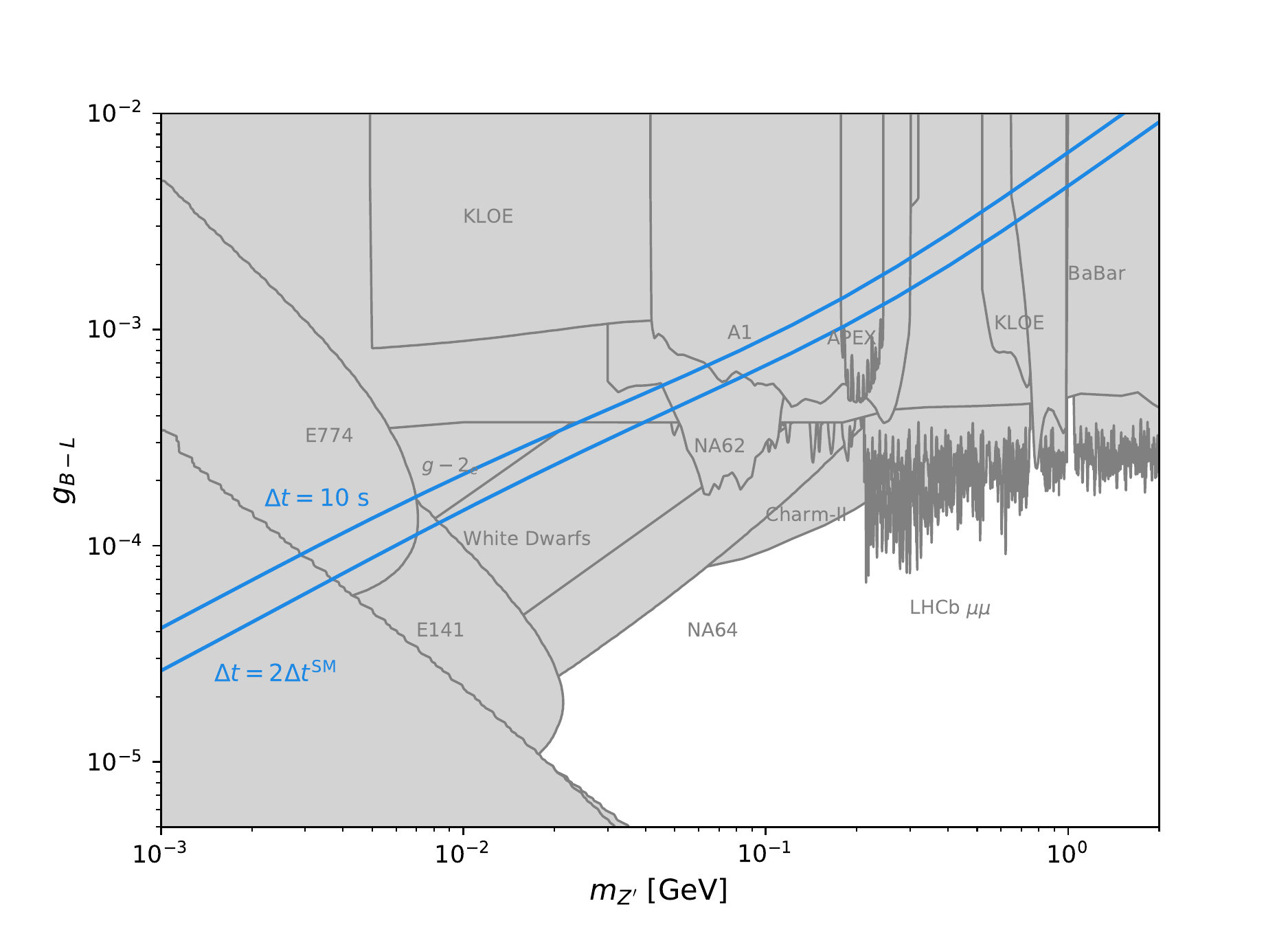}
    \vspace*{-2ex}
    \caption{As in \cref{fig:LNV_constraints,fig:LNC_constraints} but for a $U(1)_{B-L}$ vector mediator. In this model there is only one independent coupling, $g_{B-L}$, so constraints from different sources can be compared directly.}
    \label{fig:vector_constraints}
\end{figure}

Using the values of the mean free path with medium effects, $\lambda_k$, we compute the neutrino diffusion time for each point in the the parameter space of our low-mass mediator models through \cref{eq:deltat}, and we determine the regions where the diffusion conditions of \cref{eq:diff1,eq:diff2} are satisfied. We represent the results in Figs.\,\ref{fig:LNV_constraints} and \ref{fig:LNC_constraints} for a LNV and LNC scalar mediator, respectively, and in \cref{fig:vector_constraints} for the \ubl scenario. 
Since the main contribution to the neutrino-matter interaction rate comes from scattering off neutrons, the relevant cross section is proportional to the product of the coupling of neutrinos to the new mediator and the coupling of the mediator to neutrons. Thus, in the scalar model, we place constraints on $Y=\sqrt{C_\nu C_N}$. For the \ubl, where there is only one independent coupling, we choose $g_{B-L}$.

There are numerous experiments which have probed different aspects of new physics in the neutrino sector (or new light mediators in general), without having found so far any significant deviation with respect to the SM predictions. This has lead to constraints on different combinations of couplings in the new physics model. In the \ubl model, all of these constraints can be translated directly into bounds on $g_{B-L}$. We take our preexisting constraints on this model from Refs. \cite{Amaral:2020tga,Bauer:2018onh}. In the two scalar models, we consider three classes of constraints, divided based on the combination of couplings they apply to. In \cref{fig:LNC_constraints,fig:LNV_constraints}, we plot all of the constraints under the assumption of universal couplings to SM fermions ($C_q=C_l=C_\nu=Y/\sqrt{13.8}$), though they can be rescaled for other models with different relative couplings.

\newpage

\begin{itemize}
    \item \textbf{Neutrino-nucleus interactions.}
    
    The constraints we derive from supernova neutrino scattering can only be directly compared with other limits on $Y=\sqrt{C_\nu C_N}$. This is the same combination of couplings that appears in the cross section for \coherent. This elusive process has been measured by two experiments within the COHERENT collaboration, the first time in CsI \cite{Akimov:2017ade}, and again more recently in the CENNS-10 liquid argon experiment \cite{Akimov:2020pdx}. Both results were compatible with the SM prediction, and can therefore be used to place bounds on modifications to the \coherent scattering rate. Constraints on a light scalar mediator were first derived in Ref.~\cite{Farzan:2018gtr,Akimov:2018vzs}, and were later improved using an updated quenching factor \cite{Papoulias:2019txv,Khan:2019cvi}. They are identical for the LNV and LNC scalar models.

    Constraints can also be obtained on $Y$ by combining limits on the individual couplings $C_N$ and $C_\nu$. The former have been derived from measurements of neutron scattering \cite{Pokotilovski:2006up}, while the latter can be obtained from cosmology and, in the case of the LNV scalar, from searches for neutrinoless double-beta decay. Cosmological bounds arise from limits on the effective number of neutrino degrees of freedom during Big-Bang nucleosynthesis (BBN), and differ for the LNV and LNC models. All of these are discussed in more detail in Ref.~\cite{Farzan:2018gtr}.  
    
    We represent these constraints with solid lines in \cref{fig:LNC_constraints,fig:LNV_constraints}.
    
    \item {\bf Neutrino-electron interactions.}
    
    Constraints can be derived on the combination of couplings $\sqrt{C_\nu C_e}$ from measurements of neutrino-electron scattering. In \cref{fig:LNC_constraints,fig:LNV_constraints}, we use dot-dashed lines to represent the equivalent bounds that would apply to $Y$ for the case where $C_e=C_q$. These bounds can be adapted for a theory with a different relation between these couplings, by rescaling them by a factor $C_q/C_e$.

    We have obtained constraints from the results of the Borexino experiment \cite{Agostini:2017ixy}, following the method used in Ref.~\cite{Amaral:2020tga}. We have also included the limits that can be derived from the GEMMA detector, using the results of Ref~\cite{Boehm:2020ltd}, and from the recent analysis of electron recoil events at XENON1T \cite{Aprile:2020tmw,Boehm:2020ltd,AristizabalSierra:2020edu,Amaral:2020tga}. In both cases these constraints are less relevant than those from Borexino.

    \item {\bf Quark-lepton interactions.}
    
    Constraints have also been placed on models with new light mediators from the decays of mesons containing heavy quarks. In Ref.~\cite{Aaij:2016qsm}, measurements of branching ratios of $B^+$-meson decays were used to place limits on a new scalar mixing with the SM Higgs. The constraints in that work were obtained due to the effective coupling generated between a top quark (in a loop) and a $\mu^+ \mu^-$ pair. This constraint can be rescaled to give a constraint on the combination of couplings $\sqrt{C_t C_\mu}$ in our scalar models. The resulting constraints on $Y$ are shown as dotted lines in \cref{fig:LNC_constraints,fig:LNV_constraints} under the assumption of universal couplings to SM fermions.

    As with the neutrino-electron scattering bounds, the constraints derived from $B^+$-meson decays only apply to $Y$ in a model-dependent way, and should be rescaled with the relative $C_t$ and $C_\mu$ couplings in a scalar model with non-universal couplings. However, even in the case where there is no tree-level coupling to top quarks, a constraint could still be obtained by replacing the top quark in the loop with either its first- or second-generation counterpart. Such a scenario would require more careful consideration than the simple rescaling used for the neutrino-electron scattering constraints.
    
\end{itemize}

In \cref{fig:LNV_constraints,fig:LNC_constraints}, and \cref{fig:vector_constraints} we show in solid lines and under the labels $\Delta t =10$ s and $\Delta t =2 \Delta t^{\rm SM}$ the upper bounds based on the diffusion time constraints given by \cref{eq:diff1,eq:diff2} for the LNV (blue and orange) and LNC (blue) scalar mediator and for the \ubl model (blue). In Fig.\,\ref{fig:LNV_constraints} the orange lines correspond to the case in which $E_{\nu}^M=\mu_\nu^D+\pi T$ while the blue ones correspond to $E_{\nu}^M=\pi T$. As it has been already shown in Section \ref{sec:diff}, the SM diffusion time is smaller for lower values of $E_{\nu}$. This translates into a bigger gap between the $\Delta t =10$~s and $\Delta t =2 \Delta t^{\rm SM}$ lines when $E_{\nu}^M=\pi T$ (as aforementioned, the realistic neutrino energy for Majorana neutrinos should be obtained through numerical simulations). The differences between the diffusion upper bounds for the LNV, \cref{fig:LNV_constraints}, and LNC, \cref{fig:LNC_constraints}, scenarios for the same incoming neutrino energy, stem from the fact that medium effects are different for Majorana and Dirac neutrinos since in the first case $\mu_\nu=0$. Note that these constraints have been weakened with respect to the ones of~\cite{Farzan:2018gtr}, mainly due to the effect of the Pauli blocking which restricts the phase space of nucleons (for LNV and LNC) and neutrinos (only for LNC). As a result of this, SN diffusion constraints are no longer competitive with those of COHERENT.

Regarding the \ubl model, the bounds of Fig.~\ref{fig:vector_constraints} are also less restrictive than other existing limits. The different behaviour of these bounds with respect to the LNC scalar scenario, especially at low mediator masses, is due to the distinct scattering amplitudes, \cref{eq:Msq_scalar,eq:MBL}, associated to each model. Note that in these cases the nature of the neutrino is the same (Dirac) and medium effects are introduced in an identical way.

For the LNV and the LNC scenarios, we also show in \cref{fig:LNV_constraints,fig:LNC_constraints} the EoS limit given by \cref{eq:eos}, and the region (shaded in blue) where the conversion of $\nu_L$ into $\nu_R$ for the LNC scenario could distort the neutrino burst flux and time. As already mentioned above, the supernova EoS line is only indicative and shows the region above which LNV interactions could affect the EoS of SN matter. Likewise, the excluded region due to $\nu_R$ production in the LNC scenario is based on qualitative arguments (the upper limit corresponds to demanding at least $\sim 100$ new physics interactions in \cref{eq:nu_R_trapping} to consider that $\nu_R$ is trapped).
This region has shifted upwards with respect to the results of Ref.~\cite{Farzan:2018gtr} due to Pauli blocking and other medium effects, disfavouring a narrow band that was allowed in their work for $Y\sim10^{-4}$.

Care must be taken when comparing our constraints with previous results in the literature \cite{Farzan:2018gtr, Suliga:2020jfa}, as there are substantial differences in the analysis. 
In particular, none of these previous works have included medium effects, i.e., Pauli blocking for the outgoing states and effective masses, and chemical potentials for nucleons. As we have shown in \cref{fig:medium}, these are responsible for a significant reduction in the scattering cross section (an increase in the mean-free path) which, in turn, results in a less stringent bound. The recent analysis of Ref.\,\cite{Suliga:2020jfa} also considered a radial dependence of the density in the proto-NS, however their input data corresponds to a snapshot at $\Delta t =0.25$\,s after bounce (whereas we take $\Delta t =1,\,5$\,s for consistency) and it is based on a simulation that employs a different nuclear matter model (the SFHO \cite{Steiner:2012rk}) than the one we take, where the neutrinosphere has a larger radius, $R=40$\,km. Besides, it must be noticed that their computation of the neutrino mean-free-path ignores the central region, where the nuclear density is higher, and, therefore, where neutrinos get to spend more time. This translates into weaker constraints since the region where more scatterings take place is not considered to calculate the time which neutrinos spend streaming out.

Finally, our choice of couplings for the scalar case ($Y$) coincides with the notation used in Ref.\,\cite{Farzan:2018gtr}, but the comparison with Ref.\,\cite{Suliga:2020jfa} must include a rescaling of their couplings by a factor $\sqrt{Q'}$, with $Q'\approx 14$ the nucleon coherence factor. For the $U(1)_{B-L}$ model, we can directly compare with Fig.~9 of Ref.\,\cite{Suliga:2020jfa}.
Taking this into account, we notice that our bounds both for the $U(1)_{B-L}$ model and the scalar scenario are approximately a factor $2$ more stringent than those of Ref.~\cite{Suliga:2020jfa}. This mostly comes from the fact that they neglect the inner part of the star where more scatterings take place.

\section{Conclusions}
\label{sec:conclusions}

In this article, we have reevaluated the constraints on particle models with new low-mass scalar and vector mediators in the neutrino sector that can be derived from neutrino diffusion SN. To do this, we have computed the neutrino mean free path for three simplified scenarios, featuring either a new scalar (in a LNV and LNC model) or a new vector field (in a \ubl construction), incorporating medium effects in the determination of the neutrino-nucleon scattering cross-section, and a radial dependence of the density, energy, and temperature inside the proto-NS.

The resulting diffusion time has been compared to the neutrino flux observed from SN1987A, which suggests that neutrinos do not remain trapped for longer than approximately $10$~s. Using this as an upper bound, we have derived constraints on the properties of the new low-mass mediators (namely their coupling strength to neutrinos and their mass). We have compared these bounds with those from other experimental techniques.

Our results improve previous estimations, which did not take into account matter effects. In particular, we have found that matter effects lead to an increase of the neutrino mean free path that is more prominent in the central regions of the SN core where the temperature and density are larger. We have also shown that this effect is more important when new physics contributions dominate, for which the neutrino mean free path can increase by more than one order of magnitude with respect to its value in vacuum. This, in turn, leads to shorter diffusion times and relaxes the constraints on the neutrino coupling to nucleons.

The limits derived on the scalar mediator model are less restrictive than current bounds from other experimental sources, and the bounds on coherent elastic neutrino-nucleus scattering are leading for most of the values of the mediator mass. Likewise, in the \ubl scenario, we have shown that the constraint lies comfortably within the area of the parameter space that has already been explored by neutrino experiments.

Our findings motivate dedicated numerical simulations of Majorana neutrinos in SN, which would more precisely incorporate the time and radial dependence of the stellar parameters in the computation of the neutrino diffusion time. This would also permit to study properly the effect of changes in the EoS. In line with previous works, we have indicated when new LNV interactions might alter the EoS, but we have been unable to check whether this has any effect on the SN observables.

\section*{acknowledgements}

We are grateful for conversations with Yasaman Farzan, Patrick Foldenauer and Luca Mantani. DGC acknowledges financial support from the Comunidad Aut\'onoma de Madrid through the grant SI2/PBG/2020-00005, and is supported in part by the Spanish Agencia Estatal de Investigaci\'on through the grants PGC2018-095161-B-I00 and IFT Centro de Excelencia Severo Ochoa SEV-2016-0597. The work of MC was funded by the F.R.S.-FNRS through the MISU convention F.6001.19. MAPG acknowledges financial support from Junta de Castilla y Le\'on through the grant SA096P20, Agencia Estatal de Investigaci\'on through the grant PID2019-107778GB-100  and PHAROS COST Action CA16214.
We also acknowledge support of the Spanish Consolider MultiDark FPA2017-90566-REDC. We also acknowledge use of the TITAN cluster at the Universidad de Salamanca (USAL) and the Hydra cluster at the Instituto de F\'\i sica Te\'orica (IFT), on which the numerical computations for this paper took place.

\bibliographystyle{JHEP-cerdeno}
\bibliography{biblio}

\end{document}